\documentclass[pre,reprint]{revtex4-1}
\usepackage{bm,epsfig,graphicx,color,amssymb,amsmath}

\begin{document}

\title{Excitonic effects in 2D semiconductors: Path Integral Monte Carlo approach}

\author{Kirill A. Velizhanin}
\email{kirill@lanl.gov}
\affiliation{Theoretical Division, Los Alamos National Laboratory, Los Alamos, New Mexico 87545, USA}

\author{Avadh Saxena}
\email{avadh@lanl.gov}
\affiliation{Theoretical Division, Los Alamos National Laboratory, Los Alamos, New Mexico 87545, USA}

\begin{abstract}

One of the most striking features of novel 2D semiconductors (e.g., transition metal dichalcogenide monolayers or phosphorene) is a strong Coulomb interaction between charge carriers resulting in large excitonic effects. In particular, this leads to the formation of multi-carrier bound states upon photoexcitation (e.g., excitons, trions and biexcitons), which could remain stable at near-room temperatures and contribute significantly to optical properties of such materials. In the present work we have used the Path Integral Monte Carlo methodology to numerically study properties of multi-carrier bound states in 2D semiconductors. Specifically, we have accurately investigated and tabulated the dependence of single exciton, trion and biexciton binding energies on the strength of dielectric screening, including the limiting cases of very strong and very weak screening. The results of this work are potentially useful in the analysis of experimental data and benchmarking of theoretical and computational models.

\end{abstract}

\maketitle

\section{Introduction}\label{sec:introduction}

Last decade has witnessed an ever-growing interest of the scientific
and engineering communities in atomically thin two-dimensional (2D)
materials \cite{Bonaccorso-2015-41,Butler-2013-2898,Mas-Balleste-2011-20,Miro-2014-6537}.
This is due to both the unique properties of such materials from the basic
science perspective \cite{Semenoff-1984-2449,CastroNeto-2006-1},
as well as their tremendous technological potential \cite{Bonaccorso-2015-41,Xia-2014-899,Jariwala-2014-1102,Britnel-2013-1311}.
While graphene - a flagship of this class of materials - has attracted
most of the attention to date, it lacks an electronic bandgap and
is thus unsuitable ``as is'' for multiple applications including
photovoltaics and electronics \cite{Novoselov-2012-192,Jariwala-2014-1102}.
The realization of this fact has led to significant efforts spent during the 
last few years on searching for ``graphene with a bandgap'', i.e.,
atomically thin \emph{semiconductor} materials. Of many materials
identified as potential semiconductor alternatives to graphene, one
of the most promising classes of materials is monolayer transition metal dichalcogenides
(ML-TMDC) \cite{Mak-2010-136805,Splendiani-2010-1271}. Other
possible candidates include phosphorene \cite{Liu-2014-4033,Li-2014-372,Prada-2015-245421,Seixas-2015-115437}, stanene \cite{Xu-2013-136804,Balendhran-2015-640} and ultrathin organic-inorganic perovskite crystals \cite{Yaffe-2015-045414}.

A prominent feature of 2D semiconductors (2DS) is the strong Coulomb
interaction between charge carriers, as compared to their parent
3D materials (e.g., bulk TMDC). This effect arises from the so called
\emph{dielectric confinement} - the dielectric screening of the Coulomb
interaction between two point charges within a 2D material becomes weak at distances larger than some effective thickness of the material \cite{Keldysh-1979-658,Ye-2014-214,Chernikov-2014-076802,Berkelbach-2013-045318}.
In other words, the strength of dielectric screening becomes distance-dependent,
i.e., spatially \emph{non-local}. For example, this effect is pronounced in ML-TMDC, leading to the formation of tightly-bound excitons with
binding energies up to $\sim0.4-0.7\:{\rm eV}$ \cite{Ramasubramaniam-2012-115409,Berkelbach-2013-045318,Chernikov-2014-076802,Ugeda-2014-1091,Ye-2014-214,He-2014-026803}.
This is very different from conventional 3D semiconductors, where
the optical response at room temperature is dominated by independent
charge carriers \cite{YuCardona-Fundamentals-1999}. Such large exciton
binding energies in 2DS suggest the possibility of
existence of other stable bound states consisting of a larger number
of charge carriers (e.g., trions, biexcitons) at near-room temperatures \cite{Yao-2015-448}.
Indeed, trions - bound states of a single exciton and an extra
charge carrier - with binding energies on the order of $\sim20-45\:{\rm meV}$ in ML-TMDC and up to $\sim 190\:{\rm meV}$ in few-layer phosphorene have been
recently observed \cite{Ross-2013-1474,Mak-2013-207,Mitioglu-2013-245403,Lin-2014-5569,Zhang-2015-1411.6124}.
Very recent evidence of stable bound states of two electrons and two
holes - biexcitons - with binding energies of $\sim45-70\:{\rm meV}$ in ML-TMDC has been reported in Refs.~\cite{Mai-2014-202,Shang-2015-647,You-2015-477}.

The term ``excitonic effects" has been traditionally applied to features of the optical properties of semiconductors, which go beyond the simple picture of non-interacting charge carriers. In particular, the Coulomb interaction between carriers leads  to the formation of multi-carrier bound states out of free charge carriers. Theoretically, excitonic effects have mostly been studied at the level of a single exciton using electronic structure theory tools \cite{Cheiwchanchamnangij-2012-205302,Ramasubramaniam-2012-115409,Qiu-2013-216805,Shi-2013-155304,Klots-2014-6608,Ugeda-2014-1091,Ye-2014-214,Choi-2015-066403}. Experimentally observed large exciton binding energies were largely reproduced and the detailed structure of excitonic states was studied. In particular, it has been demonstrated that even without any external dielectric screening (e.g., due to a solvent or substrate) the size of an exciton in ML-TMDC is much larger than the size of a single unit cell \cite{Qiu-2013-216805,Ugeda-2014-1091,Ye-2014-214}. Under these conditions, quasimomenta of interacting electrons and holes are distributed very close to the positions of the respective band edge extrema in the Brillouin zone, see for example the inset of Fig.~3(b) in Ref.~\cite{Qiu-2013-216805}. This observation justifies the use of the effective mass approximation in the analysis of excitonic effects in ML-TMDC \cite{Ramasubramaniam-2012-115409,Berkelbach-2013-045318,Chernikov-2014-076802,You-2015-477}.   

Despite enormous interest, however, theoretical and computational studies of multi-carrier bound states beyond a
single exciton (e.g., trions, biexcitons) in 2DS are rather sparse to date. To the best of our knowledge, the first theoretical
analysis of single excitons and trions in novel 2DS explicitly taking into account the spatially \emph{non-local} character of dielectric
screening was performed in Ref.~\cite{Berkelbach-2013-045318}, followed by Refs.~\cite{Berghauser-2014-125309,Zhang-2014-205436,Chernikov-2014-076802}.
In particular, trion binding energies were found by means of a simple few-parameter variational wave function \cite{Berkelbach-2013-045318,Zhang-2014-205436}.
Trion excitations in these systems have also been studied by means
of the fractional dimensional space approach \cite{Thilagam-2014-053523},
and the time-dependent density-matrix functional theory \cite{Ramirez-Rorres-2014-085419},
but no non-local screening effects were explicitly taken into account in these works.
Finally, an accurate analysis of single exciton and trion binding energies in the limit of a very strong non-local 2D screening has been reported very recently \cite{Ganchev-2015-107401}.

The goal of the present work is to accurately study the excitonic effects in 2DS at the \emph{arbitrary} strength
of the non-local 2D screening. Within the effective mass
approximation, we demonstrate how a general problem of finding the
energy of a bound state of a specific number of charge carriers can
be reduced to finding an unknown function of two parameters: the electron-to-hole
mass ratio and the screening length. We then evaluate
this function numerically using the Path Integral Monte Carlo (PIMC)
methodology \cite{Ceperley-1995-279}. This specific flavor of the
broad class of quantum Monte Carlo methods is among the most general
methods to study interacting quantum many-body systems. It does not
require the assumption of positivity of a wave function (diffusion
quantum Monte Carlo) \cite{Kosztin-1996-633}, or choosing a wave function
ansatz (variational quantum Monte Carlo) \cite{Ronnow-2011-035316},
and has been successfully applied to analyze the electronic structure
of low-dimensional semiconductors \cite{Borrmann-2001-3120,Filinov-2003-1441,Filinov-2006-4421,McDonald-2012-125310}. In particular, binding energies of excitons, trions and biexcitons in GaAs/AlGaAs quantum wells have been accurately evaluated \cite{Filinov-2003-1441}.

Using PIMC, we evaluate the exciton, trion and biexciton binding energies
for a range of values of the screening length assuming identical isotropic electron and hole effective masses. This constitutes the main result of the paper, since a straightforward interpolation
and rescaling of the obtained numerical results (compiled in Table~\ref{tab:energies}) can be used to evaluate binding energies of single excitons, trions and biexcitons in \emph{arbitrary} 2DS with not too different isotropic effective masses (e.g., ML-TMDC or stanene). To the best of our knowledge, this work is  the first to numerically exactly treat trions and biexcitons in 2DS, albeit within the effective mass approximation. Numerical results, obtained here, can be directly used to analyze experimental data and benchmark other theoretical and computational approaches to excitonic effects in 2DS. 

The paper is organized as follows. The general Hamiltonian of the system of interacting charge carriers is introduced in Sec.~\ref{sec:gtheory}. The dependence of the energy of a multi-carrier bound state on the electron to hole effective mass ratio and the strength of 2D screening is discussed in Sec.~\ref{sec:gprop}. Brief description of the PIMC technique and its application to finding energies of multi-carrier bound states is given in Sec.~\ref{sec:PIMC}. Section~\ref{sec:res} contains the discussion of the numerical results. An example of how these numerical results could be applied in a realistic situation is given in Section~\ref{sec:example}. Section~\ref{sec:conclusion} contains the conclusion.

\section{General Theory}\label{sec:gtheory}

Within the effective mass approximation, the Hamiltonian for an arbitrary
number of interacting charge carriers (electrons and holes) inside
a 2D semiconductor (2DS) material can be written as \cite{Keldysh-1979-658,Cudazzo-2011-085406,Berkelbach-2013-045318}
\begin{equation}
H_{\delta}=\frac{1}{2}\sideset{}{^{\delta}}\sum_{i}\frac{p_{i}^{2}}{m_{i}}+\sideset{}{^{\delta}}\sum_{i,j<i}\frac{q_{i}q_{j}}{\kappa r_0}V\left(\frac{x_{ij}}{r_0}\right),\label{eq:genH}
\end{equation}
where $p_{i}^{2}=-\hbar^{2}\left[\frac{\partial^{2}}{\partial x_i^{2}}+\frac{\partial^{2}}{\partial y_i^{2}}\right]$
and $m_{i}$ are the isotropic effective masses. We assume only two types of
charge carriers with effective masses $m_{e}$ (electrons) and $m_{h}$ (holes)
and respective charges $q_{e}=-e$ and $q_{h}=e$, where $e=|e|$ is
the magnitude of the electron charge. Number of carriers of each type
is encoded by $\delta$. In this work, we restrict ourselves to $\delta=X$,
$T_{+}$, $T_{-}$ or $XX$, which corresponds to a single exciton
($e+h$), positive trion ($e+2h$), negative trion ($2e+h$), or biexciton
($2e+2h$), respectively. Accordingly, $\sideset{}{^{\delta}}\sum$ encodes a summation only over carriers specified by $\delta$.

In this work, the energies of single carriers are defined relative to corresponding band edges, so that the 2DS bandgap does not explicitly enter Eq.~(\ref{eq:genH}). Consequently, the obtained energies of multi-carrier bound states are {\em negative}, and, for example, the energy of a single exciton obtained from Eq.~(\ref{eq:genH}) is not the energy of an exciton peak directly observed in a photoluminescence experiment. Instead, such a negative energy corresponds to an energy release upon the formation of a multi-carrier bound state out of free charge carriers.

The non-locally screened Coulomb potential in Eq.~(\ref{eq:genH}) is given by
\begin{equation}
V(x)=\frac{\pi}{2}\left[H_{0}(x)-Y_{0}(x)\right],\label{eq:H0V0}
\end{equation}
where $H_{0}(x)$ and $Y_{0}(x)$ are the Struve function and the
Bessel function of the second kind, respectively \cite{AbramowitzStegun1965}.
The effective dielectric constant of the environment is defined as
$\kappa=(\kappa_{1}+\kappa_{2})/2$, where $\kappa_{1}$ and $\kappa_{2}$
are the dielectric constants of two materials that the 2DS is surrounded by.
Screening length is denoted by $r_0$. In the case of a thin semiconductor
layer of a finite thickness $d$ and an isotropic dielectric constant
$\kappa_0$, the screening length is evaluated as $r_0=d\kappa_0/2\kappa$
\cite{Keldysh-1979-658}. In the case of a truly two-dimensional material
(e.g., atomically thin), the screening length is given by \cite{Cudazzo-2011-085406}
\begin{equation}
r_0=2\pi\chi_{2D}/\kappa,\label{eq:rs_chi}
\end{equation}
where $\chi_{2D}$ is the 2D polarizability of the material.
Asymptotic expansions of $V(x)$ at large and small $x$ are
\begin{equation}
V(x)\approx1/x
\end{equation}
and 
\begin{equation}
V(x)\approx\log(2/x)-\gamma,\label{eq:V_log}
\end{equation}
respectively, where $\gamma=0.57721...$ is Euler's constant \cite{Keldysh-1979-658,Cudazzo-2011-085406}.
The former limiting case corresponds to the Coulomb interaction unaffected
by the dielectric polarization of 2DS as most of the electric field
lines between two distant charges go outside of the 2DS. In the
opposite limit, $x_{ij}\ll r_0$, most of the electric field lines
are confined within the 2DS. In this case, the Coulomb potential becomes logarithmic. 

The dimensional analysis of Hamiltonian~(\ref{eq:genH}) with only
a single type of carrier for each charge sign results in the following
universal expression for the ground state energy
\begin{equation}
E_{\delta}=-2E_{0}\mathcal{E}_{\delta}\left(\sigma,\tilde{r}_0\right),\label{eq:Eg_univ}
\end{equation}
where $\mathcal{E}_{\delta}$ is some yet unknown dimensionless function of two
continuous real arguments, that does also depend on the number of electrons
and holes in a system via the parameter $\delta=X$, $T_{+}$, $T_{-}$
and $XX$. The arguments are $\sigma=m_{e}/m_{h}$ - ratio of electron and hole effective
masses, and $\tilde{r}_0=r_0/a$ - screening length normalized
by the Bohr radius. The Bohr radius is introduced as $a=\frac{\hbar^{2}\kappa}{\mu e^{2}}$,
where $\mu=(m_{e}^{-1}+m_{h}^{-1})^{-1}$ is the reduced mass. The Bohr
energy is $E_{0}=\frac{e^{2}}{\kappa a}$.

\section{General properties of $\mathcal{E}_{\delta}(\sigma,\tilde{r}_0)$}\label{sec:gprop}

General properties and the asymptotic behavior of $\mathcal{E}_{\delta}(\sigma,\tilde{r}_0)$
as a function of mass ratio $\sigma$ and the dimensionless 2D screening
length $\tilde{r}_0$ can be understood without performing numerical
calculations.

\subsection{\label{sub:sigma_dep}Dependence on $\sigma$}

First of all, it is important to note that since the Hamiltonian (\ref{eq:genH}) does possess the charge conjugation symmetry
($C$-symmetry), it is sufficient to specify $\mathcal{E}_{\delta}(\sigma,\tilde{r}_0)$
only in the range of $\sigma\in(0,1)$. Indeed, due to the $C$-symmetry
one always has
\begin{equation}
\mathcal{E}_{\delta}(\sigma,\tilde{r}_0)\equiv\mathcal{E}_{\bar{\delta}}(1/\sigma,\tilde{r}_0),\label{eq:c-symm}
\end{equation}
where $\bar{\delta}$ specifies a multi-carrier state obtained from
state $\delta$ by charge conjugation. For example, if $\delta=T_{+}$,
then $\bar{\delta}=T_{-}$. Therefore, knowing the energy of $T_{+}$
at $\sigma<1$ immediately gives the energy of $T_{-}$ at $\sigma>1$
and vice versa. Substitution $\sigma\rightarrow1/\sigma$ naturally
arises from the charge conjugation of $\sigma=m_{e}/m_{h}$. Two important
consequences of Eq.~(\ref{eq:c-symm}) are as follows. The first
rather trivial one is that energies of positive and negative trions
are the same at $\sigma=1$. Henceforth, we will use designation
$\delta=T$ for trions when electron and hole masses are identical
and therefore the overall charge of a trion is irrelevant for energetics.

The second consequence is that
\begin{equation}
\left.\frac{\partial}{\partial\sigma}\mathcal{E}_{\delta}(\sigma,\tilde{r}_0)\right|_{\sigma=1}=-\left.\frac{\partial}{\partial\sigma}\mathcal{E}_{\bar{\delta}}(\sigma,\tilde{r}_0)\right|_{\sigma=1},\label{eq:derv_sigma}
\end{equation}
which can be directly obtained by differentiation of both sides of
Eq.~(\ref{eq:c-symm}). For trions, this equation means that if, for
example, the energy of the positive trion decreases with $\sigma$
at fixed $\mu$ and $\sigma\sim1$, then the energy of the negative
trion necessarily \emph{increases} with the same magnitude of the derivative.
For charge-neutral multi-carrier states, i.e., excitons and biexcitons
in this work, we have $\delta\equiv\bar{\delta}$, and, therefore
\begin{equation}
\left.\frac{\partial}{\partial\sigma}\mathcal{E}_{\delta}(\sigma,\tilde{r}_0)\right|_{\sigma=1}=0.\label{eq:der_neutral}
\end{equation}
Since the energy of a single exciton does not depend on the electron
or hole effective mass independently, but only through the reduced
mass $\mu$, $\mathcal{E}_{X}(\sigma,\tilde{r}_0)$
does not actually depend on $\sigma$, so that Eq.~(\ref{eq:der_neutral})
is redundant for $\delta=X$. However, $\mathcal{E}_{\delta}(\sigma,\tilde{r}_0)$
does in general depend on the mass ratio for bound states of more
than two carriers. Consequently, Eq.~(\ref{eq:der_neutral}) suggests
a somewhat non-trivial result for biexcitons - weak dependence of
$\mathcal{E}_{XX}(\sigma,\tilde{r}_0)$ on $\sigma$ when electron
and hole effective masses are similar.

In the opposite limit, i.e., when the mass ratio $\sigma$ becomes very small, one has $m_{e}\approx\mu$ and $m_{h}\approx\mu/\sigma\gg m_{e}$. Under these conditions, the Born-Oppenheimer approximation can be applied by treating holes as classical particles.
Then, the first quantum-mechanical correction to the relative motion
of holes would be to treat that motion as vibrational normal modes
of an ``electron-hole molecule'' with frequencies $\omega\sim m_{h}^{-1/2}$.
The zero-point energy of such vibrational normal modes leads to a
non-analytic asymptotic behavior of $\mathcal{E}_{\delta}(\sigma,\tilde{r}_0)$
at $\sigma\rightarrow0$
\begin{equation}
\mathcal{E}_{\delta}(0,\tilde{r}_0)-\mathcal{E}_{\delta}(\sigma,\tilde{r}_0)\propto\sigma^{1/2}.\label{eq:zpe_corr}
\end{equation}
These considerations, however, require the presence of more than a
single hole. Therefore, we expect Eq.~(\ref{eq:zpe_corr}) to be
accurate for $T_{+}$ and $XX$ in this work. For $T_{-}$, we expect
a more regular behavior at $\sigma\rightarrow0$.

\subsection{Dependence on $\tilde{r}_0$}

In this subsection, we consider the behavior of $\mathcal{E}_{\delta}(\sigma,\tilde{r}_0)$
in the limit of very weak ($\tilde{r}_0\rightarrow0$) and very
strong ($\tilde{r}_0\rightarrow\infty$) 2D screening. In the former
limit, the screening becomes strictly local as it originates only
from the 3D environment (via $\kappa$) and not from 2DS. In this
limit, the exact analytical solution of the single exciton problem
exists yielding $\left.\mathcal{E}_{X}(\sigma,\tilde{r}_0)\right|_{\tilde{r}_0\rightarrow0}=1$
\cite{YuCardona-Fundamentals-1999}. Hence the choice of the prefactor
of $2$ in Eq.~(\ref{eq:Eg_univ}). In the case of a trion or biexciton,
the exact analytical expression for $\mathcal{E}_{\delta}(\sigma,0)$
is not available, but multiple approximate analytical, as well as
approximate numerical and numerically exact results have been obtained
to date \cite{Bressanini-1998-4956,Liu-1998-588,Ronnow-2011-035316,Sergeev-2001-597,Sergeev-2005-541,Shiau-2012-115210,Shiau-2013-309,Singh-1996-15909,Stebe-1989-545,Thilagam-1997-7804,Thilagam-2014-053523,Usukura-1999-5652}. 

In the opposite limit of a very strong 2D screening, $\tilde{r}_0\rightarrow\infty$,
the Coulomb potential becomes logarithmic [see Eq.~(\ref{eq:V_log})],
so that Hamiltonian (\ref{eq:genH}) can be rewritten as 
\begin{equation}
H_{\delta}=\frac{e^{2}}{\kappa a}\left[-\frac{1}{2}\sideset{}{^{\delta}}\sum_{i}\frac{1}{\tilde{m}_{i}}\Delta_{\tilde{x}_{i}}+\sideset{}{^{\delta}}\sum_{i,j<i}\frac{\tilde{q}_{i}\tilde{q}_{j}}{\tilde{r}_0}\left(\log\frac{2\tilde{r}_0}{\tilde{x}_{ij}}-\gamma\right)\right],\label{eq:Hlog_org}
\end{equation}
where all the variables inside the square brackets are made dimensionless:
$\tilde{m}_{i}=m_{i}/\mu$, $\tilde{x}_{i}=x_{i}/a$, $\tilde{q}_{i}=q_{i}/e=\pm1$.
We further rescale coordinates by introducing $\xi_{i}=\tilde{x}_{i}/\tilde{r}_0^{1/2}$,
which yields
\begin{equation}
H_{\delta}=\frac{e^{2}}{\kappa r_0}h_{\delta}+E_{\delta}^{s},\label{eq:Hlog}
\end{equation}
where
\begin{equation}
h_{\delta}=-\frac{1}{2}\sideset{}{^{\delta}}\sum_{i}\frac{1}{\tilde{m}_{i}}\Delta_{\xi_{i}}-\sideset{}{^{\delta}}\sum_{i,j<i}\tilde{q}_{i}\tilde{q}_{j}\log\xi_{ij},\label{eq:Hxi}
\end{equation}
and
\begin{equation}
E_{\delta}^{s}=-\frac{e^{2}}{\kappa r_0}\frac{N_{e}+N_{h}-(N_{e}-N_{h})^{2}}{2}\left(\frac{1}{2}\log4\tilde{r}_0-\gamma\right),
\end{equation}
where $N_e$ and $N_h$ are the numbers of electrons and holes, respectively, within a multi-carrier bound state. Since the effective masses in atomic units can be expressed as $\tilde{m}_{e}=1+\sigma$
and $\tilde{m}_{h}=1+\sigma^{-1}$, Hamiltonian in Eq.~(\ref{eq:Hxi})
depends only on $\sigma$ and $\delta$. To the best of our knowledge,
Hamiltonians of this type were previously studied for a single exciton
\cite{Doman1982-863,Keldysh-1979-658}, and very recently for a trion
\cite{Ganchev-2015-107401}, but not for a biexciton or any other
bound state of more than three carriers. Denoting the lowest eigenvalue
of $h_{\delta}(\sigma)$ in Eq.~(\ref{eq:Hxi}) as $f_{\delta}(\sigma)$,
we obtain the general result for an arbitrary number of carriers
\begin{align}
\mathcal{E}_{\delta}\left(\sigma,\tilde{r}_0\right)&=-\frac{f_{\delta}(\sigma)}{2\tilde{r}_0}\nonumber \\
&+\frac{N_{e}+N_{h}-(N_{e}-N_{h})^{2}}{4\tilde{r}_0}\left(\frac{1}{2}\log4\tilde{r}_0-\gamma\right).\label{eq:E_xi}
\end{align}
This expression suggests that by finding a single eigenvalue of Hamiltonian
(\ref{eq:Hxi}) for specific $\delta$ and $\sigma$, one can recover
the entire dependence of $\mathcal{E}_{\delta}(\sigma,\tilde{r}_0)$
on $\tilde{r}_0$ in the limit of strong 2D screening \cite{Keldysh-1979-658,Doman1982-863,Ganchev-2015-107401}. 

Equation~(\ref{eq:E_xi}) becomes very simple in the case of a single exciton
\begin{equation}
\mathcal{E}_X(\sigma,\tilde{r}_0)=\frac{\log 4\tilde{r}_0-2f_X(\sigma)-2\gamma}{4\tilde{r}_0}.\label{eq:assy_exc}
\end{equation}
This expression has a ``built in" applicability range in the sense that it becomes unphysically negative at $\tilde{r}_0\lesssim 1$ if we assume $f_X(\sigma)\sim 1$.

\section{Path Integral Monte Carlo}\label{sec:PIMC}

In this section we briefly outline the general ideas behind the Path
Integral Monte Carlo (PIMC) methodology and specific details of its
application to finding many-body eigenstates of Hamiltonian (\ref{eq:genH}).
Most of the static properties of a finite-temperature system defined
by a Hamiltonian operator $H$ can be obtained from a generating functional
(partition function) \cite{Tsvelik-2003-Quantum} 
\begin{equation}
Z={\rm Tr}\left[e^{-\beta H}\right],\label{eq:ZV}
\end{equation}
where $\beta=1/k_{B}T$, and ${\rm Tr}$ represents the quantum mechanical trace. Functional derivatives of the so defined partition
function with respect to $H$ can then be used to evaluate most of
the relevant observables of the system. For example, the charge density
of the system can be obtained by introducing an auxiliary electrostatic
potential to $H$ and then finding the functional derivative of $Z$
with respect to this potential.

The quantum mechanical trace in Eq.~(\ref{eq:ZV}) can be represented by a discrete path integral 
\begin{equation}
Z=\int\prod_{m=1}^{N}dx^{m}\,\langle x^{m+1}|e^{-\tau H}|x^{m}\rangle,\label{eq:Z_dpi}
\end{equation}
where $\tau=\beta/N$ is an imaginary time step, and $x^{m}$ specifies
the coordinates of the entire system at the $m^{{\rm th}}$ moment
of imaginary time. Periodic boundary conditions are assumed, $x^{N+1}\equiv x^{1}$.
If the time step is small enough (i.e., approaching the limit of continuous
path integral), then the matrix elements in Eq.~(\ref{eq:Z_dpi})
can be approximated by the so called symmetrized Suzuki-Trotter decomposition
formula as \cite{Suzuki-1985-601,Ceperley-1995-279}
\begin{align}
\langle x^{m+1}|e^{-\tau H}|x^{m}\rangle&=\mathcal{C}\exp\left\{ -\frac{\sum_{i}m_{i}\left(x_{i}^{m+1}-x_{i}^{m}\right)^{2}}{2\tau}\right.\nonumber\\
& \left. -\frac{\tau}{2}\left[U(x^{m+1})+U(x^{m})\right]\right\} +\mathcal{O}(\tau^{3}),\label{eq:ST}
\end{align}
where $x_{i}^{m}$ is a coordinate of the $i^{{\rm th}}$ particle
at the $m^{{\rm th}}$ moment of the imaginary time, $\mathcal{C}$ is the normalization
constant and $U(x)$ is the potential energy part of $H$. The partition
function can now be evaluated approximately as an approximate multi-dimensional
integral representation of the continuous path integral, Eqs.~(\ref{eq:Z_dpi})
and (\ref{eq:ST}).

Various observables of the system can be obtained by finding the functional
derivative of the integrand in Eq.~(\ref{eq:Z_dpi}) with respect
to $H$ and then evaluating the resulting integral. One of the most
efficient ways to evaluate such integrals for large $N=\beta/\tau$
is to employ the Metropolis-Hastings Monte Carlo algorithm \cite{Metropolis-1953-1087,Ceperley-1995-279},
hence the name ``Path Integral Monte Carlo''.

\begin{figure}
\includegraphics[width=3.2in]{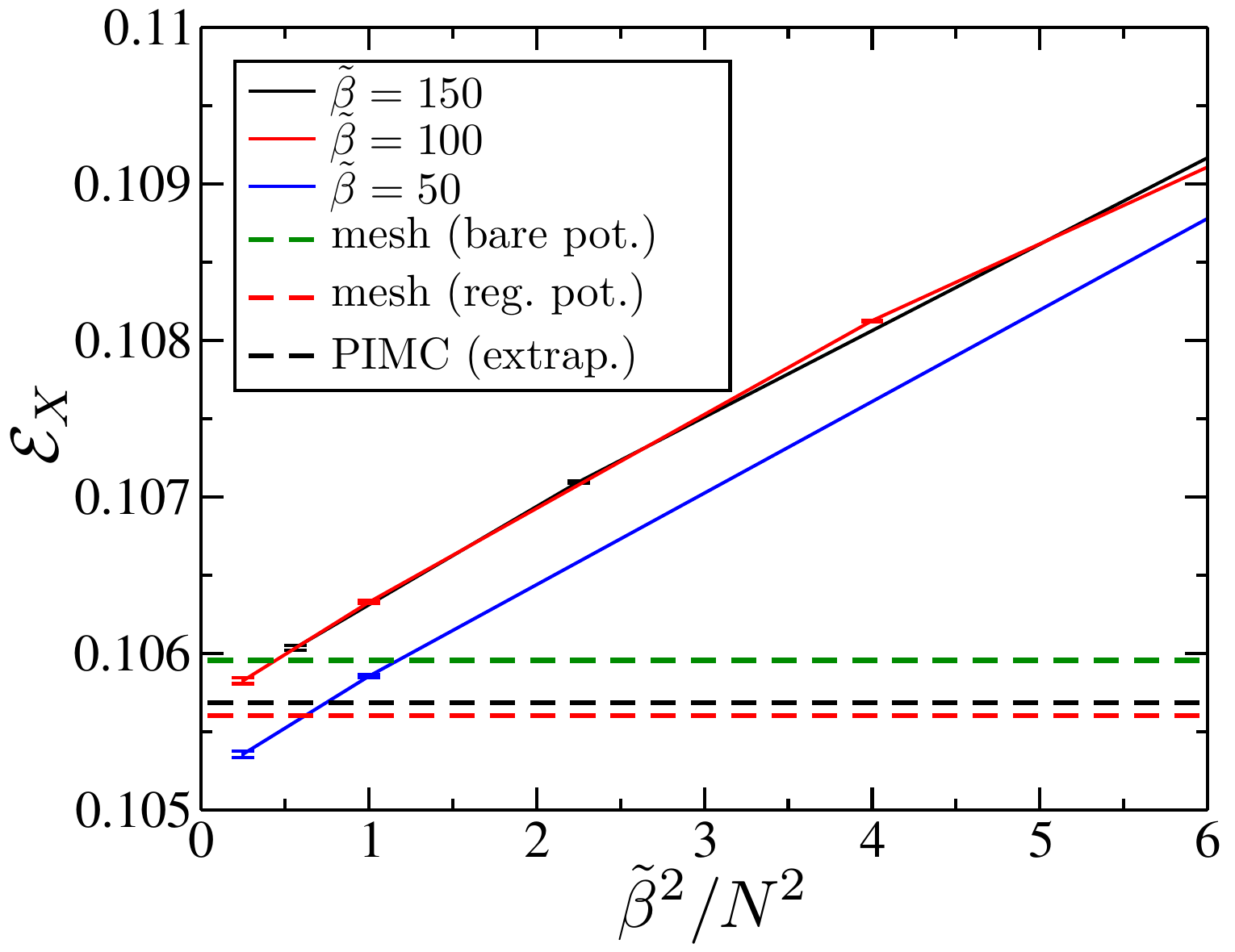}
\caption{\label{fig:converg}
PIMC results for the dimensionless exciton binding
energy $\mathcal{E}_{X}$ as a function of $\tilde{\beta}^{2}/N^{2}$
are plotted as solid lines. System parameters are $\tilde{r}_0=5$
and $\sigma=1$. The extrapolation of the solid red and black lines
to $N\rightarrow\infty$ yields the binding energy of the exciton
ground state, shown by dashed black line. Red and green horizontal
dashed lines correspond to grid-based calculations with and without
regularization of the inter-carrier interaction potential (see text
for details).}
\end{figure}

It can be demonstrated that the error of the order $\sim\tau^{3}$ in Eq.~(\ref{eq:ST})
transforms into the error of the order of $\sim\tau^{2}$ for observables.
Therefore, one can expect the following relation between the exact
value of an observable $A$ and its approximation $A_{N}$ obtained
using Eq.~(\ref{eq:ST})
\begin{equation}
A_{N}-A\propto\frac{\beta^{2}}{N^{2}}.\label{eq:ObsErr}
\end{equation}
If $A$ is temperature-independent at very low temperatures, then this expression can be used to extrapolate $A_{N}$ to $A$ by calculating
$A_{N}$ for a few values of $N$ and $\beta$ and then performing
a regression analysis. Figure~\ref{fig:converg} plots the dimensionless
binding energy of a single exciton, $\mathcal{E}_{X}$, as a function
of $\tilde{\beta}^{2}/N^{2}$, where the dimensionless inverse temperature
is introduced as $\tilde{\beta}=\beta E_{0}$, where $E_0$ is the Bohr energy introduced in Sec.~\ref{sec:gtheory}.
As expected, the three solid lines corresponding to PIMC results for
a range of $N$ and $\tilde{\beta}$ demonstrate a near-linear dependence
when plotted against $\tilde{\beta}^{2}/N^{2}$. Solid blue line corresponds to PIMC
calculations with the largest temperature, $\tilde{\beta}=50$. At
this temperature, there is a noticeable admixture of excited
exciton states resulting in the apparent lower binding energy (i.e.,
lower $\mathcal{E}_{X}$). The PIMC single exciton energies at lower
temperatures, $\tilde{\beta}=100$ and $150$, are shown by the red
and black lines, respectively. As can be seen, these two lines essentially
coincide signifying that the temperature is low enough to sample only
the ground state of the single exciton. Fitting these two lines with
Eq.~(\ref{eq:ObsErr}) produces the binding energy of a single exciton
in the limit $N\rightarrow\infty$, shown as the horizontal dashed
black line. 

To benchmark the PIMC result, we have also performed a simple
grid-based diagonalization of the single-exciton Hamiltonian. The
result is shown by the horizontal dashed red line and is seen to be
within the error bar of the PIMC result. 

It turns out that two modifications to Hamiltonian (\ref{eq:genH})
are necessary to make it suitable for PIMC calculations. The first
modification concerns the fact that PIMC calculations are always performed
at finite temperature ($\beta<+\infty$), and at \emph{any} finite
temperature the entropic contribution to the partition function favors
the complete dissociation of Coulomb-bound states. To prevent that
we have included a weak attractive harmonic potential to each pair
of charge carriers. Via the series of numerical tests we have demonstrated
that at sufficiently low temperature it is always possible to tune the strength
of this harmonic potential so that it is weak enough not to disturb
the ground state properties of multi-carrier bound states, and
strong enough to prevent dissociation.

The second modification of Hamiltonian (\ref{eq:genH}) concerns the
singularity of the Coulomb interaction at vanishing inter-particle
distances. Why this is a problem for PIMC calculations and how to
solve it in the case of unscreened Coulomb interaction by introducing
a \emph{regularized} Coulomb potential is discussed elsewhere \cite{Muser-1997-571}.
Our generalization of the proposed solution consists of two steps.
First, we perform a non-linear rescaling of the inter-particle distances
when calculating the potential energy
\begin{equation}
x_{ij}\rightarrow x'_{ij}(x_{ij})=\left[1-\exp(-x_{ij}/\lambda)\right]^{-1}x_{ij}.\label{eq:rescale}
\end{equation}
The rescaled distance does not go below $\lambda$, so the potential
energy does not diverge. For the case of unscreened Coulomb interaction,
this rescaling results in the regularized potential used in Ref.~\cite{Muser-1997-571}.
Second, one can notice that for potentials that are monotonic functions
of inter-particle distances, e.g., Eq.~(\ref{eq:H0V0}), the introduced
non-linear rescaling results in a non-vanishing perturbation of the
total energy of the system already in the first order with respect
to the difference between potential energy functions with and without
rescaling. This first-order deviation can be largely compensated by
adding a well-behaved short-range potential to the regularized Coulomb
potential, so that the integral over the entire 2D plane of the difference
between this resulting potential and the original potential vanishes.
Specifically, we substitute $V(x)$ in Eq.~(\ref{eq:H0V0}) with
$V'(x)=V(x')+A\exp(-x^{2}/2\lambda^{2})$, where $x'$
is defined by Eq.~(\ref{eq:rescale}) and the prefactor $A$ is found
by requiring that $\int_{0}^{R}xdx\,\left[V(x)-V'(x)\right]$
approaches $0$ as $R\rightarrow\infty$. Since both $x-x'(x)$
and $\exp(-x^{2}/2\lambda^{2})$ decay exponentially fast at large distances,
this integral approaches zero very quickly with $R$. This guarantees
that the low-momentum spatial Fourier harmonics of $V(x)$ and $V'(x)$
are almost the same, so that $V'(x)$ is a non-singular version
of $V(x)$ that is expected to reproduce the low-energy Coulomb scattering
correctly. So introduced $V'(x)$ is thus nothing but a \emph{local
pseudopotential} very similar to those used in atomistic quantum Monte
Carlo methods \cite{Burkatzki-2007-234105}. The grid-based calculations
with and without these two modifications to the interaction potential (i.e., 
harmonic potential and zero-distance regularization) are shown in 
Fig.~\ref{fig:converg} by dashed red and green lines,
respectively. As can be seen, the relative deviation of one from the other
is $\sim3\times10^{-3}$. In all the PIMC calculations in the present
work we kept the deviation at this level or less by adjusting $\lambda$. 

In general, the accurate treatment of the fermion statistics is required
to correctly reproduce eigenstates of Hamiltonian (\ref{eq:genH})
when more than a single charge carrier of the same type (charge) is
present. However, applications of quantum Monte Carlo methods to fermionic
systems are plagued by the so called ``fermion sign problem'' \cite{Ceperley-1995-279,Troyer-2005-170201,Li-2015-241117(R)}.
The easiest way to avoid this problem is to treat charge carriers
as \emph{distinguishable} particles and not fermions. This could be
done if carriers of the same type do possess ``flavors'', i.e.,
quantum numbers that are conserved by all the interactions in the
system. Then, if two carriers have different values of at least one
of such quantum numbers, they become effectively distinguishable.
For example, if the only present flavor is the spin, then two holes
(or two electrons) with opposite spin projections can be treated as distinguishable particles within
a positively (or negatively) charged trion \cite{Sergeev-2001-597,Berkelbach-2013-045318,Ganchev-2015-107401}.
In the case of ML-TMDC or stanene, the two two-valued flavors are the
spin and the \emph{pseudospin}, where the latter labels the two degenerate $K$ and $K'$ valleys in the Brillouin zone \cite{Xu-2014-343}. These two flavors
are independently conserved only approximately since strong excitonic effects and
valley-dependent spin-orbit coupling could in principle lead to mixing
between the spin and pseudospin projections. However, this mixing
was found to be weak due to the large momentum and high energy barrier
separations of the two valleys \cite{Xu-2014-343}. At these conditions, the 
``flavor multiplicity" is four because up to four same-charge carriers can be 
distinguishable in ML-TMDC and stanene. Even more, same-charge carriers 
can become distinguishable in lead chalcogenide semiconductors where there are 
four degenerate valleys, and, therefore, the total flavor multiplicity (i.e., including spin) is eight \cite{Dimmock-1964-A821,Mitchell-1966-581,Kang-1997-1632}.

Pauli repulsion suggests that the lowest-energy configuration of a multi-carrier bound state would consist of effectively distinguishable particles (i.e., with different spin or pseudo-spin projections), if the number of same-type particles is less than the overall flavor
multiplicity, which is four in ML-TMDC and stanene, and eight in lead chalcogenides. Since, this work is
focused on properties of lowest-energy configurations of $X$, $T_{+}$,
$T_{-}$ and $XX$, we neglect the fermion statistics and treat all charge carriers as distinguishable. 

\section{Numerical Results}\label{sec:res}

\begin{figure}
\includegraphics[width=3.2in]{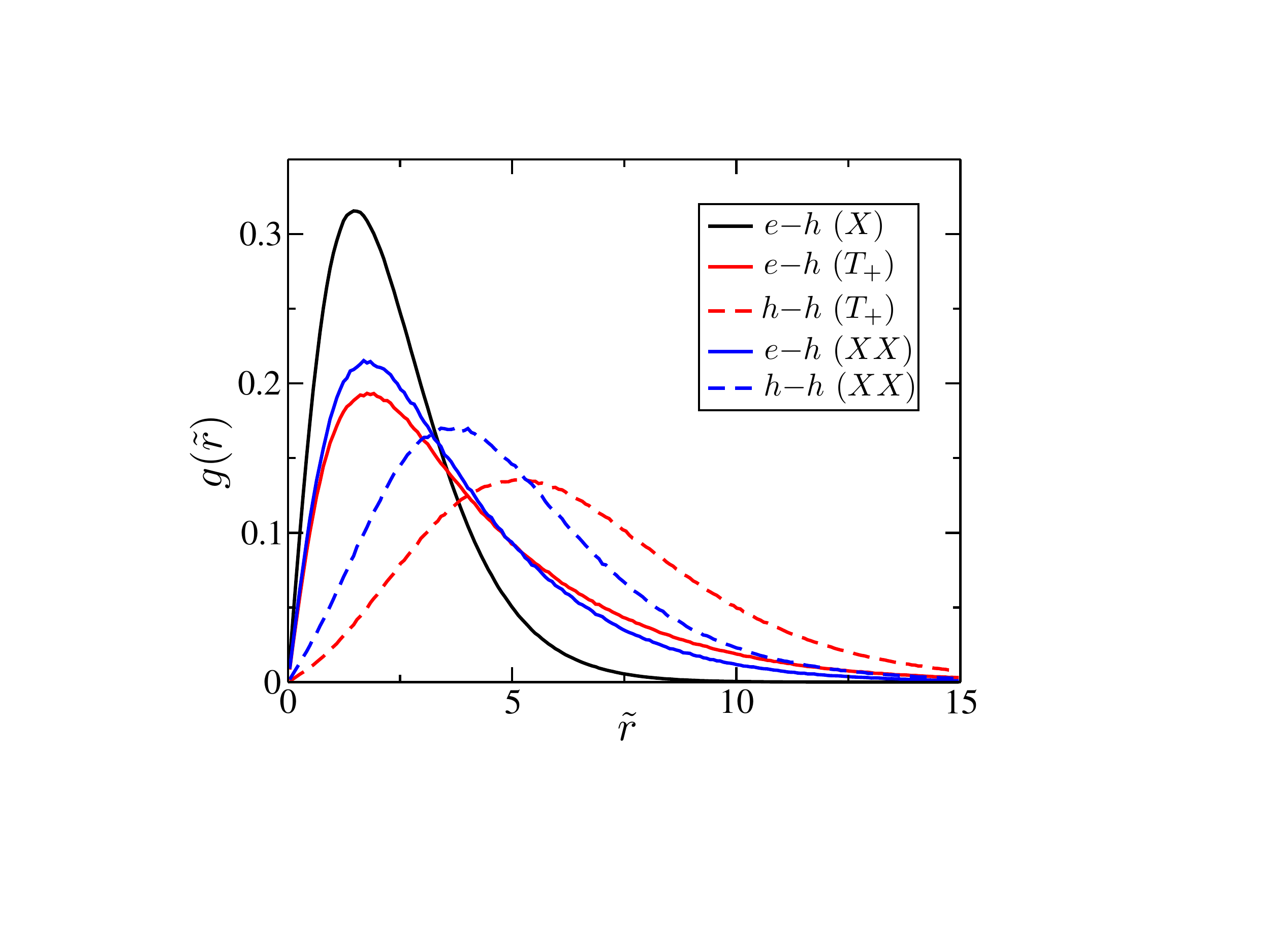}
\caption{\label{fig:rdf}
RDFs for inter-carrier distances within an exciton (black), trion (red) and biexciton (blue) for $\sigma=1$ and $\tilde{r}_0=5$. Electron-hole and hole-hole RDFs are shown in solid and dashed lines, respectively. Distance (horizontal coordinate) is given in atomic units. RDFs are normalized so that $\int d\tilde{r}\,g(\tilde{r})=1$.}
\end{figure}

PIMC-generated radial distribution functions for a single exciton,
trion and biexciton at $\sigma=1$ and $\tilde{r}_0=5$ are plotted
Fig.~\ref{fig:rdf}.
Radial distribution function (RDF) for two particles ($1$ and $2$)
is introduced as
\begin{equation}
g(r_{12})=\int\prod_{i}dx_{i}\,|\Psi(x_{1},x_{2},...,x_{N})|^{2}\delta(r_{12}-|x_{1}-x_{2}|),
\end{equation}
where $\Psi(x_{1},x_{2},...,x_{N})$ is the wave function of the ground
state of the Hamiltonian in Eq.~(\ref{eq:genH}). So introduced radial
distribution function is normalized as $\int_{0}^{\infty}d\tilde{r}_{12}\,g(\tilde{r}_{12})=1$,
and since the wave function is finite everywhere, the radial distribution
functions demonstrate the standard 2D behavior as they become linearly
proportional to $r_{12}$ at small distances.

Of all the RDFs in Fig.~\ref{fig:rdf}, the excitonic one (black
line) is seen to result in the smallest mean electron-hole distance.
This is of course expected as trions and biexcitons are supposed to
be much ``looser'' bound states with larger mean inter-carrier distances \cite{Singh-1996-15909,Usukura-1999-5652}. If we assume that $\tilde{r}_0=5$
is large enough so that the strong 2D screening limit is valid and
Eq.~(\ref{eq:Hlog_org}) is accurate, then we can estimate the excitonic
size as $\langle\tilde{r}_{X}\rangle\sim\langle\tilde{r}_{X,eh}\rangle\sim\tilde{r}_0^{1/2}\langle\xi_{X,eh}\rangle$.
Furthermore, since all the parameters in $h_{\delta}$ in Eq.~(\ref{eq:Hxi})
are on the order of $\sim1$ at $\sigma=1$, we can estimate the exciton
size in $\xi$-coordinates as $\langle\xi_{X,eh}\rangle\sim1$. This
results in $\langle\tilde{r}_{X}\rangle\sim2$, which is in a rather
good agreement with the position of the maximum of the excitonic RDF
in Fig.~\ref{fig:rdf}.

Binding energies of a trion and biexciton are conventionally introduced
as
\begin{equation}
\mathcal{E}_{T}^{b}=\mathcal{E}_{T}-\mathcal{E}_{X},\label{eq:Ebt}
\end{equation}
and
\begin{equation}
\mathcal{E}_{XX}^{b}=\mathcal{E}_{XX}-2\mathcal{E}_{X},\label{eq:Ebxx}
\end{equation}
that is the binding energy of trion is defined as the energy released
upon adding one more carrier to an exciton. The biexciton binding
energy is defined as the energy released upon the formation of a bound
state of two single initially isolated excitons. At $\tilde{r}_0\rightarrow0$,
$\mathcal{E}_{T}^{b}$ and $\mathcal{E}_{XX}^{b}$ are typically much
smaller in magnitude than the single exciton binding energy $\mathcal{E}_{X}$ ($\mathcal{E}_{X}\equiv\mathcal{E}^b_X$ in this work) \cite{Stebe-1989-545,Thilagam-1997-7804,Sergeev-2001-597,Liu-1998-588,Singh-1996-15909}.
Below we demonstrate that this qualitative picture remains true at arbitrary $\tilde{r}_0$. As a result, carriers can be thought to form
isolated excitons within the zeroth-order picture. Consequently, the first-order correction to this picture consists of accounting for
weak interactions of these neutral bound electron-hole pairs between
each other and with uncompensated charge carriers (e.g., in a trion)  \cite{Sergeev-2001-597}.
For instance, an electron in a positive trion is expected to ``see''
one of the two holes being ``excitonic'' with characteristic electron-hole
distances being comparable to the size of a single isolated exciton. The other
hole is farther away as it forms a much more weakly bound hole-exciton complex. These heuristic considerations are in agreement
with what can be deduced from the electron-hole RDFs for a trion (solid
red line) and biexciton (solid blue line) in Fig.~\ref{fig:rdf}.
Indeed, each of these two lines has two prominent features: (\emph{i})
a maximum very close to the excitonic one (solid black line), thus
representing the ``excitonic'' hole, and (\emph{ii}) a heavy shoulder
at larger distances, representing the other more weakly bound hole.
The Coulomb repulsion between carriers of the same charge results
in the maxima of the hole-hole RDFs (dashed red and blue lines) being
displaced toward larger distances, as compared to those of the electron-hole RDFs.

\begin{figure}
\includegraphics[width=3.2in]{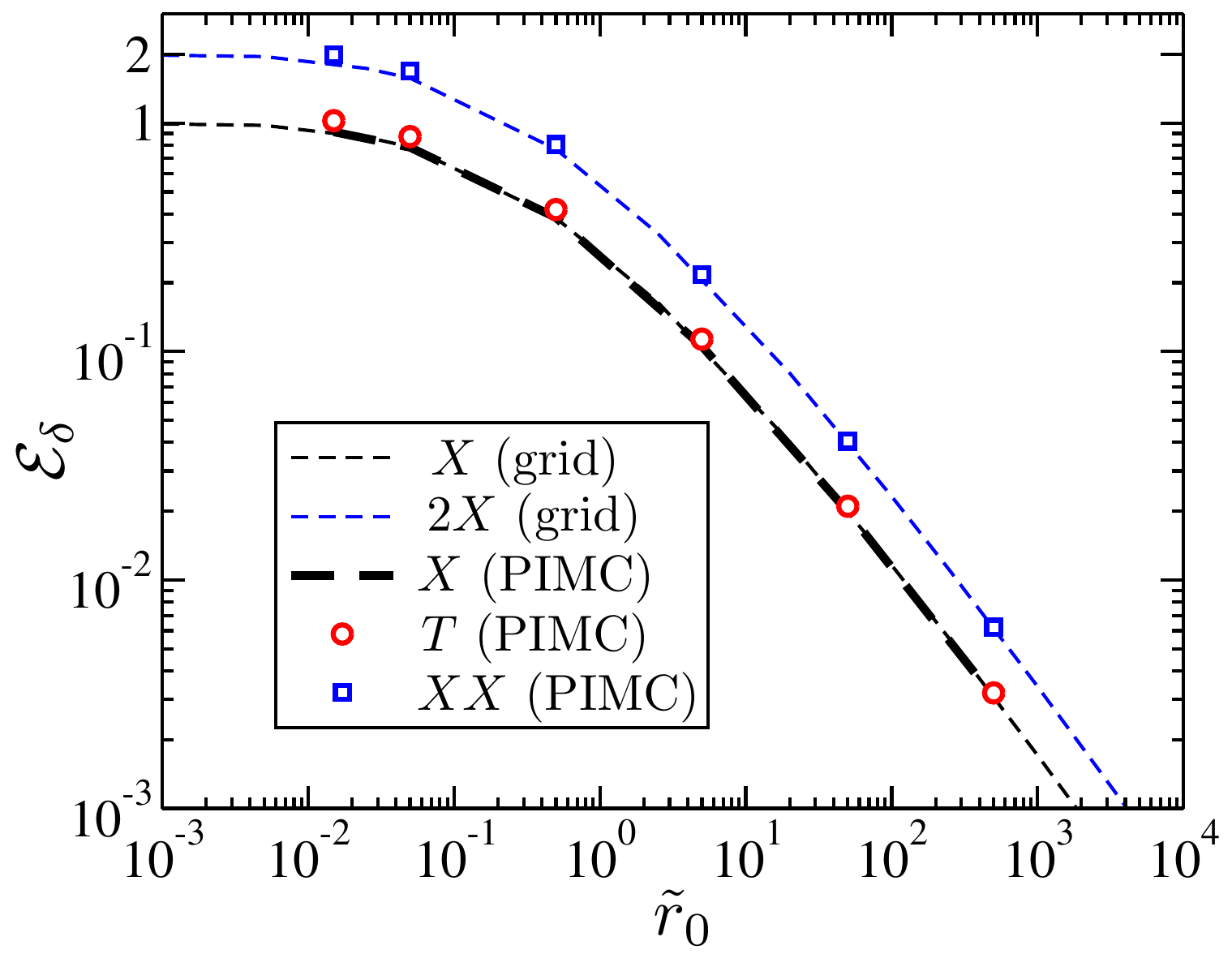}
\caption{\label{fig:E-x-t-xx}
Dimensionless energy of an exciton, trion and biexciton,
$\mathcal{E}_\delta(\sigma,\tilde{r}_0)$, as a function of $\tilde{r}_0$
at $\sigma=1$. Thin dashed black line represents the results of the grid-based
calculations of the single exciton energy. Energy of a dissociated biexciton (i.e., two isolated excitons), obtained from grid-based calculations, is shown by the thin dashed blue line. PIMC results for the exciton energy are shown by the thick dashed line. Red circles and blue
squares represent the PIMC calculations for trion and biexciton energies,
respectively.}
\end{figure}

Dimensionless energy of an exciton, trion and biexciton, $\mathcal{E}_{\delta}(\sigma,\tilde{r}_0)$,
as a function of $\tilde{r}_0$ at $\sigma=1$ is shown in Fig.~\ref{fig:E-x-t-xx}.
The results of the grid-based calculations of the single exciton energy
are given by a thin dashed black line and seen to be in excellent agreement
with the corresponding PIMC results (thick dashed black line). Red circles
and blue squares represent PIMC results for trion and biexciton energies,
respectively. As can be seen, the trion energies (red circles) are
very close to the single exciton ones (thick dashed black line). Similarly, biexciton
energies (blue squares) are very close to the energies of two isolated excitons (thin blue dashed line). PIMC results in Fig.~\ref{fig:E-x-t-xx} constitute the main result of this paper since, with the appropriate rescaling using Eq.~(\ref{eq:Eg_univ}), they can be straightforwardly
used to evaluate the exciton, trion and biexciton energies for an \emph{arbitrary}
2D semiconductor if the electron and the hole effective masses are
isotropic and not too different from each other. To facilitate the
use of this data in analysis of experimental results or benchmarking
of theoretical models, we also provide this same data in Table~\ref{tab:energies}.
\begin{table}
\begin{tabular}{|c|c|c|c|c|c|c|}
\hline 
$\tilde{r}_0$ & $\mathcal{E}_{X}$ & $\sigma_{X}$ & $\mathcal{E}_{T}$ & $\sigma_{T}$ & $\mathcal{E}_{XX}$ & $\sigma_{XX}$\tabularnewline
\hline 
0 & 1 & - & 1.125 & - & 2.1928 & -\tabularnewline
\hline 
0.015 & 0.914 & 2e-3 & 1.024 & 3e-3 & 1.993 & 5e-3\tabularnewline
\hline 
0.05 & 0.783 & 2e-3 & 0.873 & 3e-3 & 1.692 & 6e-3\tabularnewline
\hline 
0.5 & 0.385 & 1e-3 & 0.4184 & 2e-4 & 0.805 & 1e-3\tabularnewline
\hline 
5 & 0.10569 & 2e-5 & 0.1132 & 8e-5 & 0.2169 & 6e-5\tabularnewline
\hline 
50 & 1.992e-2 & 2e-5 & 2.104e-2 & 1e-5 & 4.049e-2 & 2e-5\tabularnewline
\hline 
500 & 3.072e-3 & 3e-6 & 3.203e-3 & 2e-6 & 6.213e-3 & 2e-6\tabularnewline
\hline 
\end{tabular}\protect\caption{\label{tab:energies}
PIMC results for dimensionless energies of single
excitons ($\mathcal{E}_{X}$), trions ($\mathcal{E}_{T}$) and biexcitons
($\mathcal{E}_{XX}$) calculated for a range of $\tilde{r}_0$ at $\sigma=1$. Error bars for these energies are given by
$\sigma_{X}$, $\sigma_{T}$ and $\sigma_{XX}$, respectively. The
first column is the dimensionless 2D screening length $\tilde{r}_0$. At vanishing 2D
screening ($\tilde{r}_0\rightarrow0$), the trion and biexciton
energies are obtained from Refs.~\cite{Stebe-1989-545} and \cite{Bressanini-1998-4956},
respectively. The analytical solution for the exciton binding energy at $\tilde{r}_0\rightarrow0$
is $\mathcal{E}_{X}=1$ \cite{YuCardona-Fundamentals-1999}.}
\end{table}

\begin{figure}
\includegraphics[width=3.2in]{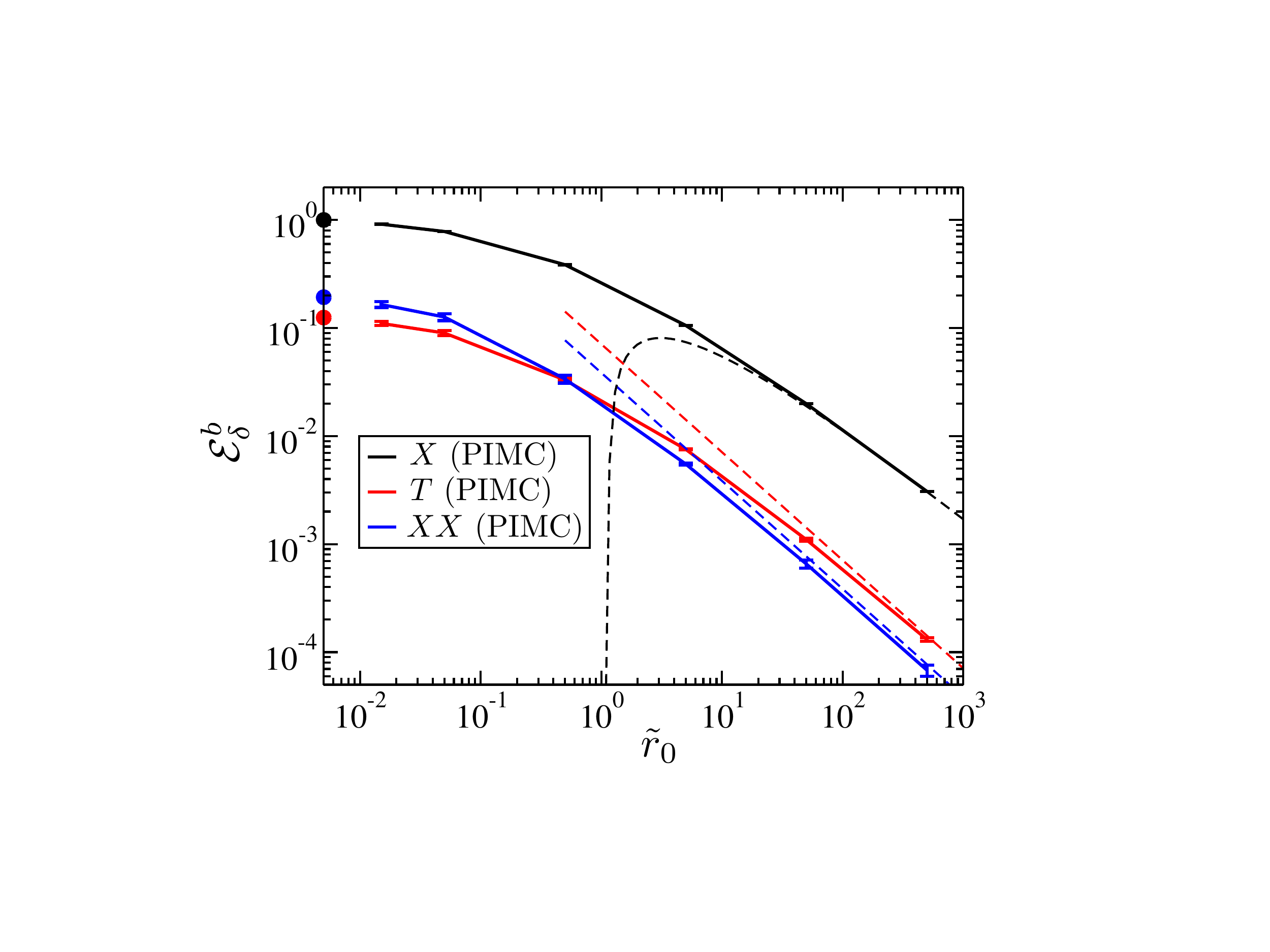}
\caption{\label{fig:Eb}
Binding energies for single excitons (black), trions (red) and biexcitons (blue) vs $\tilde{r}_0$ at $\sigma=1$. PIMC results are shown as solid lines with error bars. Large-$\tilde{r}_0$ asymptotics, Eq.~(\ref{eq:E_xi}), are shown by dashed lines. Red and blue filled circles represent the trion and biexciton binding energies at $\tilde{r}_0\rightarrow0$ obtained from Refs.~\cite{Stebe-1989-545,Bressanini-1998-4956}.}
\end{figure}

Binding energies of trions, Eq.~(\ref{eq:Ebt}), and biexcitons,
Eq.~(\ref{eq:Ebxx}), are compared to the single exciton binding
energy, $\mathcal{E}_{X}$, in Fig.~\ref{fig:Eb}. 
Specifically, binding energies of single excitons, trions and biexcitons
as a function of $\tilde{r}_0$ (at $\sigma=1$) are depicted by
black, red and blue solid lines, respectively. The lowest 2D screening length for which we have performed PIMC calculations is $\tilde{r}_0=0.015$.
PIMC calculations with even smaller $\tilde{r}_0$ become increasingly
difficult because the Coulomb potential becomes more singular at vanishing inter-particle
distances \cite{Muser-1997-571}. Fortunately, the numerical results for the trion and biexciton
binding energy at exactly $\tilde{r}_0=0$ were obtained previously by
other methods \cite{Stebe-1989-545,Bressanini-1998-4956}. These energies
are depicted in Fig.~\ref{fig:Eb} by red and blue filled circles,
respectively. The analytical solution for the 2D exciton, $\mathcal{E}_{X}(\sigma,\tilde{r}_0)=1$,
is depicted by a black filled circle. As can be seen, our numerical results
for a few smallest $\tilde{r}_0$ come quite close to the $\tilde{r}_0=0$
limit, so that a relatively smooth interpolation is possible and there
is no need for explicit PIMC calculations with very small 2D screening
lengths. 

At very large $\tilde{r}_0$, Eq.~(\ref{eq:E_xi}) is expected
to become accurate at describing energies of multi-carrier bound
states in 2D semiconductor materials. We performed PIMC calculations
for the ground states of $h_{\delta}$, Eq.~(\ref{eq:Hxi}), for the exciton, trion and biexciton at $\sigma=1$. The numerical
results are 
\begin{gather}
f_{X}=0.1800\pm0.0005, \nonumber \\
f_{T}=0.0382\pm0.0005,\nonumber \\
f_{XX}=0.283\pm0.002.
\end{gather}
Up to numerical accuracy and notation, $f_{T}$
obtained here coincides with the one obtained recently in Ref.~\cite{Ganchev-2015-107401}.
Using these constants in Eq.~(\ref{eq:E_xi}), we plot the resulting
asymptotic dependencies of binding energies as dashed lines in Fig.~\ref{fig:Eb}.
As can be seen, a good agreement between the exact PIMC results
for the single exciton binding energy and the asymptotic dependence, Eq.~(\ref{eq:assy_exc}),
is already reached at $\tilde{r}_0\sim10$, and becomes
progressively better at larger $\tilde{r}_0$. On the other hand, the asymptotic dependence deviates significantly from the exact solution at $1\lesssim\tilde{r}_0\lesssim 10$, finally becoming unphysically negative at $\tilde{r}_0\lesssim 1$.

Evaluation of the trion and biexciton binding energies in the limit of large $\tilde{r}_0$ is done via combining Eq.~(\ref{eq:E_xi}) with Eqs.~(\ref{eq:Ebt}) and (\ref{eq:Ebxx}). This produces $\mathcal{E}^b_T=\left(f_X-f_T\right)/2\tilde{r}_0$ and $\mathcal{E}^b_{XX}=\left(2f_X-f_{XX}\right)/2\tilde{r}_0$, i.e., the $f_\delta$-independent terms in Eq.~(\ref{eq:E_xi}) cancel each other out when $\mathcal{E}^b_{T}$ and $\mathcal{E}^b_{XX}$ are evaluated. Consequently, these expressions yield two straight red and blue dashed lines for trions and biexcitons, respectively, when plotted in the log-log axes in Fig.~\ref{fig:Eb}. The PIMC results for a trion and biexciton are seen to converge slower with $\tilde{r}_0$ to their corresponding large-$\tilde{r}_0$ asymptotics, than those for a single exciton. The rationale is that the inter-carrier distances within trions
and biexcitons are larger than those within single excitons (see Fig.~\ref{fig:rdf}),
so $\tilde{r}_0$ has to be larger to provide the same extent of
agreement between the finite-$\tilde{r}_0$ and asymptotic $\tilde{r}_0\rightarrow\infty$ results.

\begin{figure}
\includegraphics[width=3.2in]{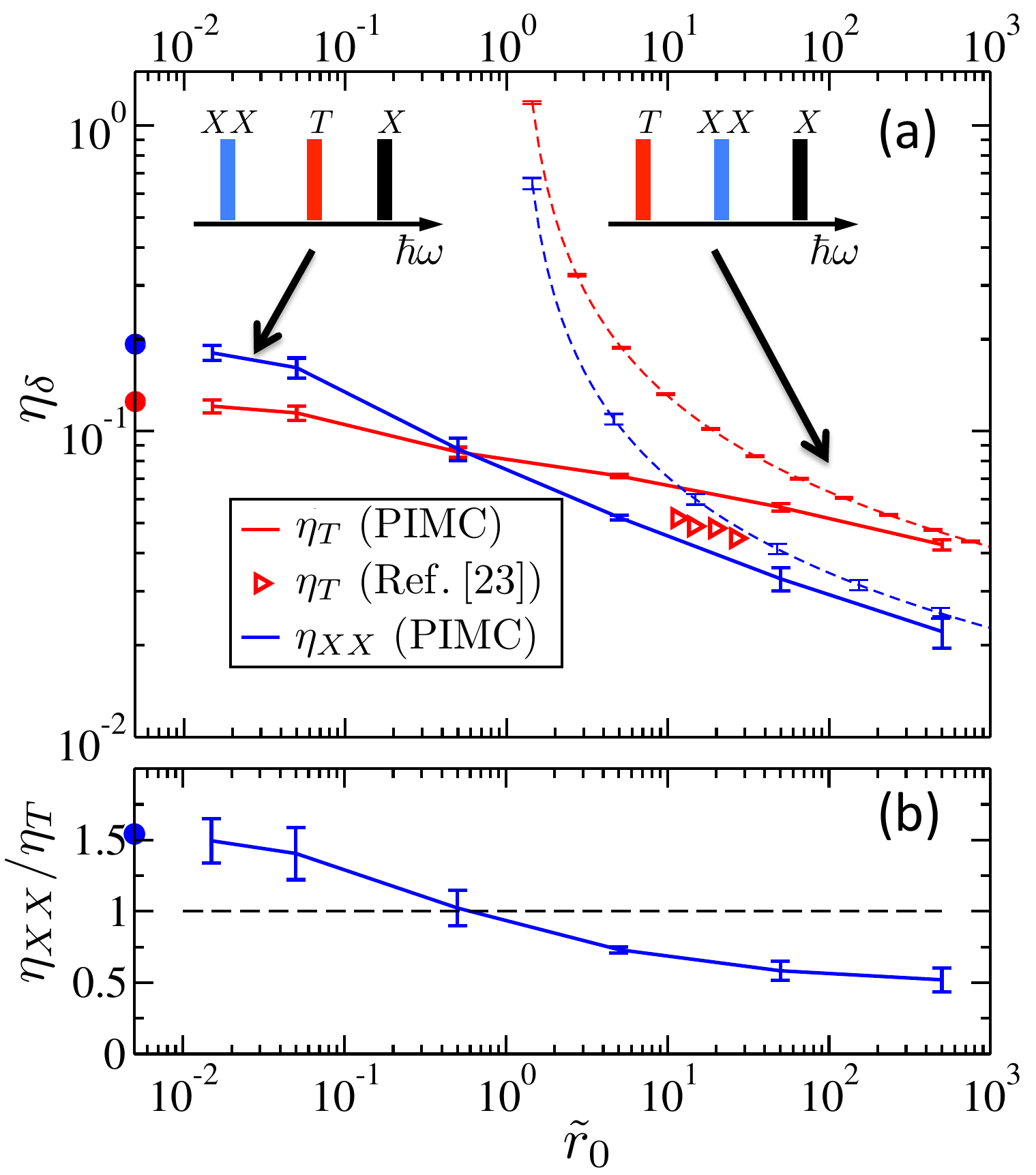}
\caption{\label{fig:tri_xx_bind}
(a) PIMC results for the Haynes factor - binding energy of a trion (solid red line) or biexciton (solid blue line) normalized by the exciton binding energy - as a function of screening length $\tilde{r}_0$ at $\sigma=1$. Dashed lines
of respective colors are the large-$\tilde{r}_0$ asymptotics, Eq.~(\ref{eq:E_xi}).
Red and blue filled circles represent the trion and biexciton Haynes factors at $\tilde{r}_0\rightarrow0$ obtained in Refs.~\cite{Stebe-1989-545,Bressanini-1998-4956}.
Red triangles are the trion Haynes factors, obtained in Ref.~\cite{Berkelbach-2013-045318} for a series of ML-TMDC materials.
Insets show schematically the energy ordering of single exciton, trion and biexciton peaks in a photoluminescence spectrum for weak (left) and strong (right) 2D screening.\newline
(b) The ratio of biexciton and trion Haynes factors as a function of screening length $\tilde{r}_0$. The black dashed line is a guide for the eye for $\eta_{XX}/\eta_T=1$. The blue filled circle represents the ratio of Haynes factors at $\tilde{r}_0\rightarrow0$ \cite{Stebe-1989-545,Bressanini-1998-4956}.
}
\end{figure}

In Fig.~\ref{fig:Eb}, the dependence of binding energies on $\tilde{r}_0$
is qualitatively the same decaying function for excitons, trions and
biexcitons except for a slowly varying prefactor seen as an almost constant
vertical shift in the log-log axes when going from single excitons to trions or biexcitons.
To focus on this prefactor it is convenient to introduce the so called trion (biexciton)
Haynes factor \cite{Haynes-1960-361} as the binding
energy of a trion (biexciton) normalized by the binding energy of a single exciton taken at the
same $\tilde{r}_0$ 
\begin{equation}
\eta_{T}=\left(\mathcal{E}_{T}-\mathcal{E}_{X}\right)/\mathcal{E}_{X},
\end{equation}
and
\begin{equation}
\eta_{XX}=(\mathcal{E}_{XX}-2\mathcal{E}_{X})/\mathcal{E}_{X}.
\end{equation}
The Haynes factors for a trion (solid red line) and a biexciton (solid blue line) are plotted in Fig.~\ref{fig:tri_xx_bind}(a) at $\sigma=1$.
Dashed red and blue lines correspond to the large-$\tilde{r}_0$
asymptotic behavior, Eq.~(\ref{eq:E_xi}), for trions and biexcitons,
respectively. Red triangles represent trion Haynes factors for a series
of four ML-TMDC materials, calculated in Ref.~\cite{Berkelbach-2013-045318}
by means of a simple variational ansatz approximation for a trion wave function. As expected, these variational results
underestimate the exact binding energy (solid red line), but by no more than $\sim20-35\:\%$.
On the other hand, it is seen that in the range of $\tilde{r}_0$ corresponding to the red triangles,
the large-$\tilde{r}_0$ asymptotics (red dashed line) can overestimate the exact trion
binding energy (red solid line) by up to a factor of $\sim2$. The dielectric screening
by the environment is expected to decrease $\tilde{r}_0$ even further
\cite{Chernikov-2014-076802}, thus potentially making the large-$\tilde{r}_0$
limit even less accurate when applied to realistic ML-TMDC.

Numerical results shown in Figs.~\ref{fig:Eb} and \ref{fig:tri_xx_bind}(a) can be used to obtain the relative
spectral positions of photoluminescence (PL) peaks pertaining to a single exciton, trion and biexciton. Indeed, if one
assumes that the final state of a radiative recombination of a trion is a single charge carrier occupying the lowest-energy band edge state, then the spectral energy gap between single exciton and trion PL peaks, $\hbar\omega_X-\hbar\omega_T$, is identical to the trion binding energy
\footnote{At finite charge doping, the final lowest-energy single carrier state is not the band edge one since the band is partially occupied. The dissociation energy, $\hbar\omega_X-\hbar\omega_T$, is then higher than the binding energy, the difference being the Fermi energy of the partially occupied band. The binding energy can then be defined as the dissociation energy in the limit of vanishing charge doping \cite{Mak-2013-207,Mitioglu-2013-245403}. Zero charge doping is assumed in this work, so the dissociation and binding energies are equal here.}.
Similarly, assuming that the final state of a biexciton radiative recombination is a single exciton in its lowest energy configuration, $\hbar\omega_X-\hbar\omega_{XX}$ is identical to the biexciton binding energy. 

Figs.~\ref{fig:Eb} and \ref{fig:tri_xx_bind}(a) suggest that the binding energy of a trion is larger than that of a biexciton at small $\tilde{r}_0$, while the opposite is true at large $\tilde{r}_0$. The crossover, i.e., where the trion and biexciton binding energies become identical, occurs at $\tilde{r}_0\approx 0.5$. This crossover is further highlighted in Fig.~\ref{fig:tri_xx_bind}(b), where the ratio of Haynes factors for a biexciton and trion is plotted as a function of $\tilde{r}_0$. 
Left and right insets in Fig.~\ref{fig:tri_xx_bind}(a) show schematically the relative position of $X$, $T$ and $XX$ PL peaks at small and large $\tilde{r}_0$, respectively. The former (left) PL spectrum is typical for semiconductor quantum wells where the effects of dielectric confinement are weak and the the dielectric screening of the Coulomb interaction is local ($\tilde{r}_0\rightarrow 0$) \cite{Filinov-2003-1441}. On the other hand, note that the biexciton binding energy can in principle be smaller than that of a trion, as has been demonstrated both experimentally and theoretically for semiconductor carbon nanotubes \cite{Watanabe-2012-035416,Yuma-2013-205412,Bondarev-2014-245430}.
In particular, a crossover very similar to that in Figs.~\ref{fig:Eb} and \ref{fig:tri_xx_bind} has been theoretically observed in Ref.~\cite{Watanabe-2012-035416} for single-wall semiconductor carbon nanotubes. The crossover has been attributed to non-local screening effects, which is again very similar to what is observed in the present work.

\subsection{Energy variation with $\sigma$}

All the calculations above were performed assuming identical effective
masses of electrons and holes, i.e., $\sigma=m_{e}/m_{h}\equiv1$.
This is because the electron and hole effective masses are similar
in ML-TMDC, $\sigma\gtrsim0.5$ \cite{Cheiwchanchamnangij-2012-205302,Ramasubramaniam-2012-115409,Kormanyos-2013-045416},
which is presently one of the most important classes of 2DS. In what follows we show that the variation
of binding energies of multi-carrier bound states is typically not
very strong at $\sigma\sim1$. Under these conditions, a binding energy
at $\sigma\equiv 1$ could be a reasonable approximation for binding energies
at $\sigma\gtrsim0.5$.

\begin{figure}
\includegraphics[width=3.2in]{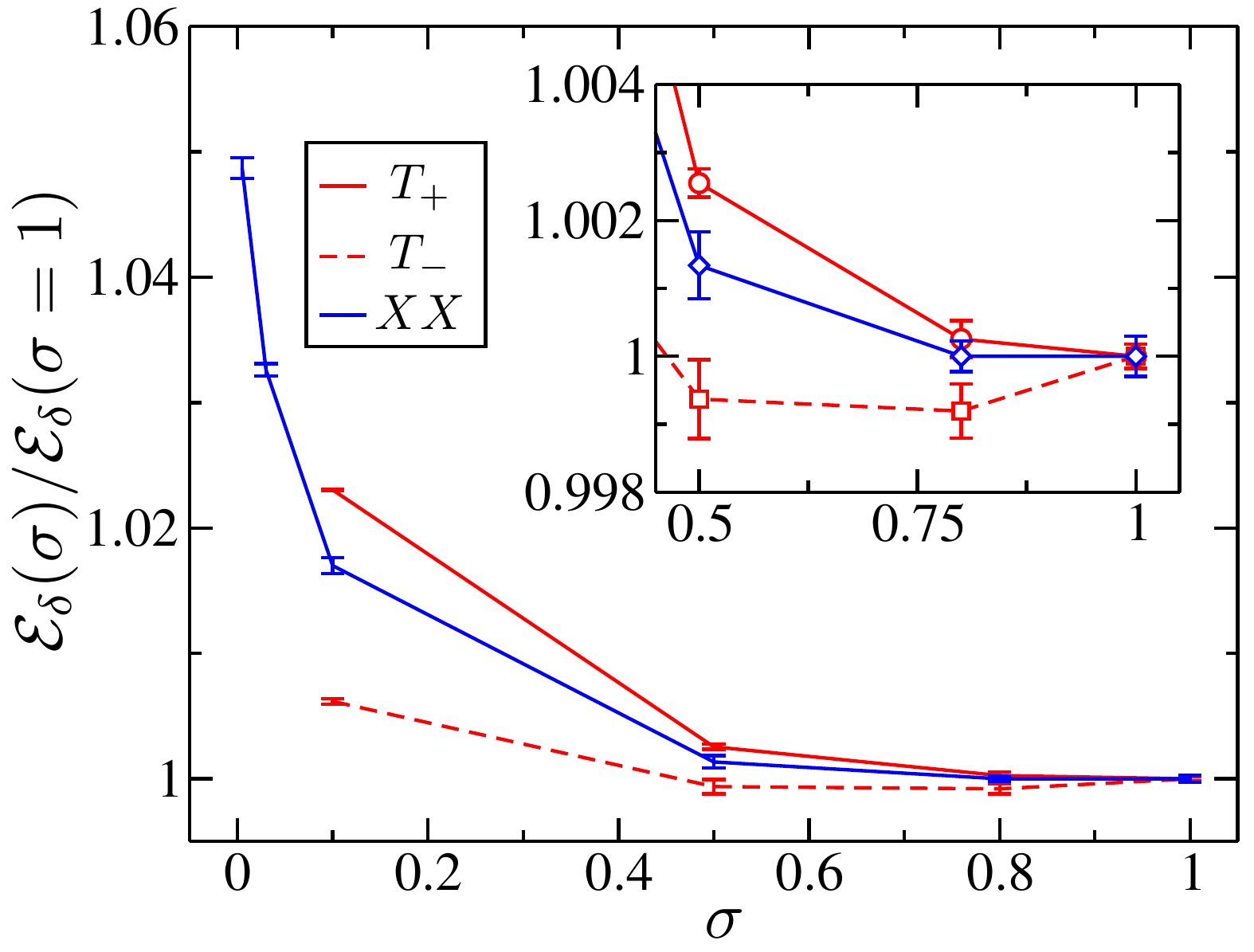}
\caption{\label{fig:sigma_dep}
The dependence of the positive trion (solid red line), negative trion (dashed red line) and biexciton (solid blue line) energies on $\sigma$ at $\tilde{r}_0=5$. The
energies $\mathcal{E}_\delta(\sigma)$ are normalized by $\mathcal{E}_\delta(\sigma=1)$. Inset shows the same data, emphasizing the weak variation of $\mathcal{E}_\delta(\sigma)$ with $\sigma$ at $\sigma\sim1$. }
\end{figure}

The weak variation of biexciton energies with $\sigma$ at $\sigma\sim1$
and arbitrary $\tilde{r}_0$ was already suggested in Sec.~\ref{sub:sigma_dep}
on the basis of vanishing derivative $\left.\partial\mathcal{E}_{XX}(\sigma,\tilde{r}_0)/\partial\sigma\right|_{\sigma=1}$.
Single exciton energy $\mathcal{E}_X$ does not depend on the mass ratio whatsoever.
That trion energies also vary weakly with $\sigma$ at $\sigma\sim1$
is not known a priori, thus calling for direct numerical testing.
Figure~\ref{fig:sigma_dep} shows the PIMC results for $\sigma$-dependences
of multi-carrier bound state energies normalized to the respective
energies at $\sigma=1$, i.e., $\mathcal{E}_{\delta}(\sigma,\tilde{r}_0)/\mathcal{E}_{\delta}(1,\tilde{r}_0)$,
$\delta=T_{+}$, $T_{-}$ and $XX$. 
The 2D screening length is set to $\tilde{r}_0=5$. As can be seen,
the deviations of the biexciton and trion energies from those at $\sigma=1$
do not exceed $\sim0.3\%$ for $\sigma\gtrsim0.5$ (see the inset). Similarly weak variations of trion energies at $\sigma\sim1$ were previously
observed in the limit of the vanishing ($\tilde{r}_0\rightarrow0$)
and very strong ($\tilde{r}_0\rightarrow\infty$) 2D screening \cite{Sergeev-2005-541,Ganchev-2015-107401}.
It is thus not unreasonable to expect a universally weak variation
of $\mathcal{E}_{\delta}(\sigma,\tilde{r}_0)$ with $\sigma$ at arbitrary $\tilde{r}_0$ if $\sigma\sim 1$. It has to be emphasized, however,
that the absolute energy of a multi-carrier bound state, Eq.~(\ref{eq:Eg_univ}),
does depend on effective masses not only through $\sigma=m_{e}/m_{h}$, but also through the reduced mass $\mu=(m_e^{-1}+m_h^{-1})^{-1}$. Consequently, the absolute
energies can vary strongly with masses even if $\sigma$ remains the
same. However, these variations are ``trivial'' in a sense that
no new numerical analysis is needed since $\mu$ enters Eq.~(\ref{eq:Eg_univ})
algebraically.

The dependence of $\mathcal{E}_{\delta}(\sigma,\tilde{r}_0)$ on
$\sigma$ becomes stronger at $\sigma<0.5$ for trions and biexcitons.
This is especially noticeable for a positive trion and a biexciton, where
two holes become very heavy at $\sigma\rightarrow0$. Under these conditions, the adiabatic Born-Oppenheimer approximation can be applied, so that
the first finite-$\sigma$ corrections to $\mathcal{E}_{XX}(\sigma=0,\tilde{r}_0)$
and $\mathcal{E}_{T_{+}}(\sigma=0,\tilde{r}_0)$ can be calculated
as a zero-point energy of vibrational motion of the two holes relative to each other, Eq.~(\ref{eq:zpe_corr}). Resulting non-analytical dependence, $\mathcal{E}_\delta(0,\tilde{r}_0)-\mathcal{E}_\delta(\sigma,\tilde{r}_0)\propto\sigma^{1/2}$, is consistent with the rapid growth of the positive trion and biexciton energies when $\sigma$ approaches zero.

\begin{figure}
\includegraphics[width=3.2in]{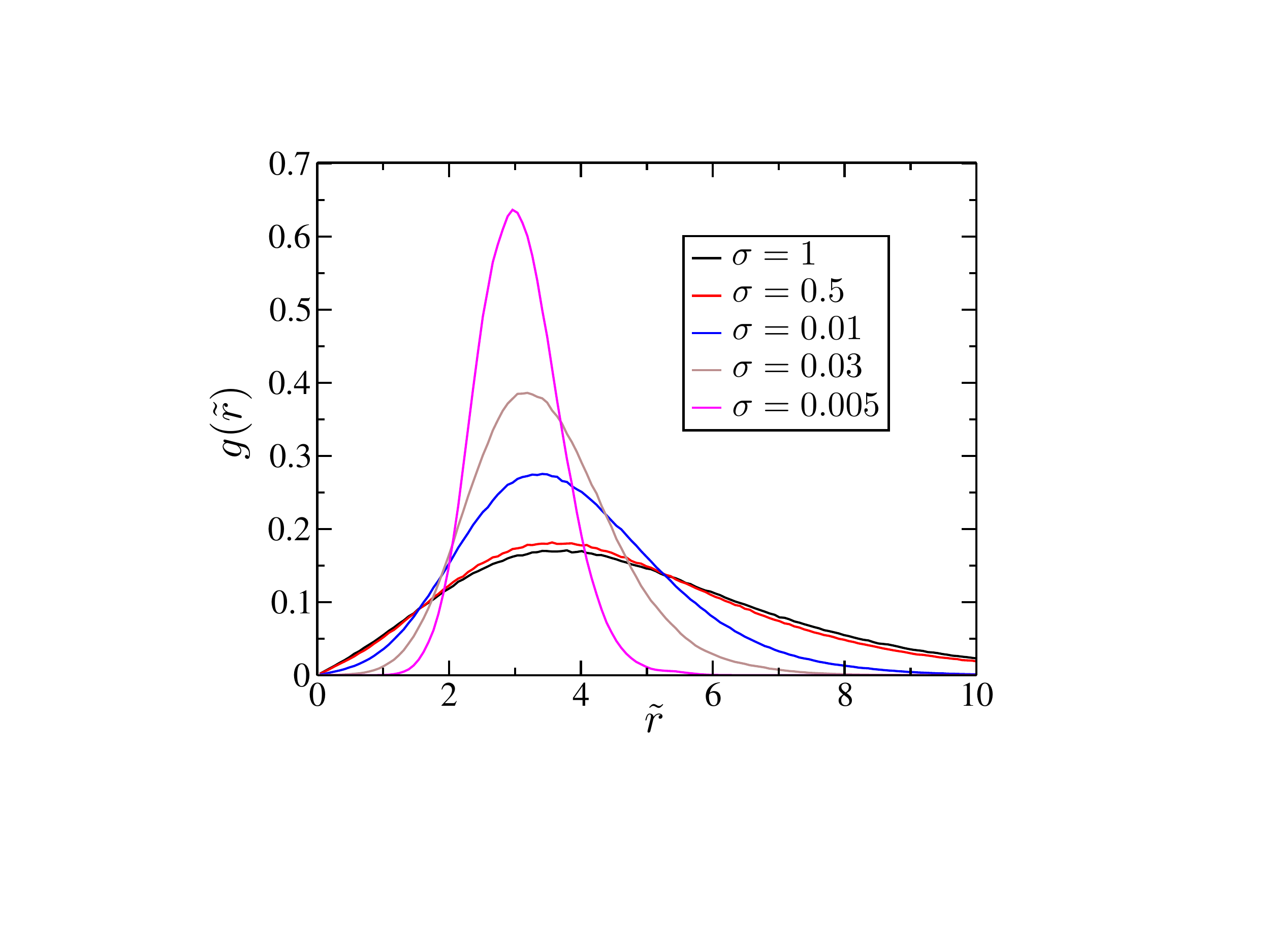}
\caption{\label{fig:rdf_sigma}
Biexciton hole-hole RDF for a range of $\sigma$ at $\tilde{r}_0=5$. }
\end{figure}

The emergent ``classicality'' of heavy holes can also be seen in
Fig.~\ref{fig:rdf_sigma} where the hole-hole RDF becomes progressively
narrower when the effective hole mass increases.
In the limit of $\sigma\rightarrow0$, the two holes would become
classical particles and, therefore, the hole-hole RDF would be a delta-function localized at finite $\tilde{r}$ at zero temperature. 

\section{Material example: ${\rm WS_2}$}\label{sec:example}

The numerical results obtained in this work can be directly used to
analyze experimental data. For convenience, we compiled all the numerically
calculated energies of multi-carrier bound states in Table~\ref{tab:energies}. Interpolating between the data points
in this table and using Eq.~\ref{eq:Eg_univ} allows one to predict binding energies of excitons,
trions and biexcitons. Other than data in the table, the following
parameters of the system are required: $\kappa$ - the effective dielectric
constant of the environment, $\chi_{2D}$ - the 2D polarizability of 2DS, as well as $m_{e}$
and $m_{h}$ - the effective masses of charge carriers. The two-dimensional polarizability
and effective masses can be obtained from the electronic structure
theory calculations \cite{Berkelbach-2013-045318,Chernikov-2014-076802}.
Some or all of these parameters can also be obtained from experiment \cite{Chernikov-2014-076802}.

To provide a concrete example, we demonstrate here how binding energies of multi-carrier bound states could be obtained for a specific material at realistic conditions - a monolayer of ${\rm WS_2}$ put on top of a ${\rm SiO_2}$ substrate. The 2D polarizability of ${\rm WS_2}$ is $\chi_{2D}=6.03\:{\rm\AA}$  \cite{Berkelbach-2013-045318}. The optical dielectric constant of the ${\rm SiO_2}$ substrate is $\kappa_{\rm SiO_2}^{\rm opt}=2.1$, resulting in the effective dielectric constant of $\kappa=(\kappa_{\rm SiO_2}^{\rm opt}+\kappa_{\rm air})/2=1.55$. Then, from Eq.~(\ref{eq:rs_chi}) we obtain $r_0\approx 24\:{\rm\AA}$. Using the exciton reduced mass $\mu=0.16$ \cite{Berkelbach-2013-045318}, we obtain the Bohr energy of $E_0\approx 1.81\:{\rm eV}$ and the Bohr radius of $a\approx 0.51\:{\rm nm}$, which results in $\tilde{r}_0\approx 4.77$. Assuming $\sigma=1$ and interpolating between the data points in Table~\ref{tab:energies} produces $\mathcal{E}_X=0.109$, $\mathcal{E}_T= 0.117$ and $\mathcal{E}_{XX}=0.224$
\footnote{Since $\mathcal{E}_X$, $\mathcal{E}_T$ and $\mathcal{E}_{XX}$ depend rather sharply on $\tilde{r}_0$, we suggest to interpolate (e.g., linearly) $\log \mathcal{E}_\delta$ as a function of $\log \tilde{r}_0$.}.
Using Eq.~(\ref{eq:Eg_univ}), these values are straightforwardly converted to exciton, trion and biexciton binding energies of 0.39 eV, 28 meV and 21 meV, respectively. Inequality of electron and hole effective masses in ${\rm WS_2}$, $\sigma\approx 0.7-0.85$ \cite{Shi-2013-155304,Kormanyos-2015-022001}, can be shown with the help of Fig.~\ref{fig:sigma_dep} to change these binding energies by at most $\pm 2\%$ of their respective values. As discussed in Sec.~\ref{sub:sigma_dep}, the exciton binding energy does not depend on $\sigma$ whatsoever, as long as the reduced mass remains fixed. The obtained exciton and trion binding energies in ${\rm WS_2}$ are consistent with the experimental values of 0.32 eV \cite{Chernikov-2014-076802} and $\sim$20-40 meV, respectively \cite{Mitioglu-2013-245403,Chernikov-2014-076802}. Changing the effective dielectric constant of the environment can strongly affect these binding energies (see e.g., Ref.~\cite{Lin-2014-5569}). For example, substitution of ${\rm SiO_2}$ with ${\rm HfO_2}$~\cite{Radisavljevic-2013-815} changes the optical dielectric constant of the substrate from $2.1$ to $\approx$4 \cite{Wood-1990-604}. The dimensionless screening length then becomes $\tilde{r}_0\approx 1.83$, yielding the exciton, trion and biexciton binding energies as 0.27 eV, 21 meV and 18 meV, respectively.


This same example reveals a peculiar contradiction between numerical results obtained in this work and very recent experimental observations of biexcitons in ML-TMDC. Specifically, the dimensionless screening length for realistic ML-TMDC-based devices can be estimated to be $\tilde{r}_0\approx 5$ \cite{Chernikov-2014-076802,You-2015-477}. Under these conditions, according to Figs.~\ref{fig:Eb} and \ref{fig:tri_xx_bind}, the ordering of PL peaks is expected to be similar to that shown schematically in the r.h.s. inset of Fig.~\ref{fig:tri_xx_bind}(a), i.e., the trion peak is expected to be lower in energy than the biexciton peak. The experimental observations suggest otherwise \cite{Shang-2015-647,You-2015-477}, i.e., the biexciton PL peak is located at lower energies than that of a trion. More quantitatively, the ratio of binding energies of a biexciton and trion is estimated from our calculations to be $\eta_{XX}/\eta_T\approx 0.74$, Fig.~\ref{fig:tri_xx_bind}(b). On the other hand, the value of $\approx1.65$ for the same ratio of binding energies can be extracted from Ref.~\cite{You-2015-477}. This latter value can only be approached with our numerical results if we assume a strictly local screening, i.e., $\tilde{r}_0\rightarrow 0$. However, such a local screening cannot explain the non-hydrogenic Rydberg series for an exciton in ML-TMDC \cite{Chernikov-2014-076802}. Provided that the experimental PL peak assignment and binding energies are correct, we need to critically rethink the applicability of the simplistic effective mass model, Eqs.~(\ref{eq:genH}) and (\ref{eq:H0V0}), for describing multi-carrier bound states in ML-TMDC. Further research in this direction is thus required.  

\section{Conclusion}\label{sec:conclusion}

In this work, using the Path Integral Monte Carlo methodology we have numerically
analyzed the problem of multi-carrier bound states in 2D semiconductors where the electron and hole effective masses are isotropic and not very dissimilar.
Specifically, we have calculated energies of single excitons, trions and biexcitons in
their lowest-energy configurations for a range of dimensionless 2D screening lengths,
$\tilde{r}_0$. These energies were demonstrated to converge correctly
to the previously analyzed limits of vanishing 2D screening (trions
and biexcitons) \cite{Stebe-1989-545,Bressanini-1998-4956}, as well
as very strong 2D screening (trions) \cite{Ganchev-2015-107401}. The latter reference could actually be considered complementary to the present work, since with the help of Eq.~(\ref{eq:Eg_univ}) the entire problem can be reformulated in terms of an unknown function of two independent parameters: the mass ratio, $\sigma=m_e/m_h$, and the dimensionless 2D screening parameter, $\tilde{r}_0$. In the present work we have systematically investigated the dependence of this function on the second parameter assuming $\sigma\sim 1$. The dependence on the first parameter was investigated in Ref.~\cite{Ganchev-2015-107401} in the limit of very large $\tilde{r}_0$.

Two limitations of the PIMC methodology, as implemented in the present work, are (i) the effective mass approximation, and (ii) its inability to accurately access energies of {\em excited states} of multi-carrier bound states. The effective mass approximation is typically justified if an exciton is of the Wannier-Mott type, i.e., the average distance between the charge carriers is much larger than the size of the unit cell of the semiconductor material. As discussed in Introduction, this is satisfied for ML-TMDC. In general, however, if an exciton is so strongly bound that the effective mass approximation fails, Bethe-Salpeter-like methods have to be used to assess excitonic effects \cite{Ramasubramaniam-2012-115409,Qiu-2013-216805,Shi-2013-155304,Ugeda-2014-1091,Ye-2014-214,Choi-2015-066403}.

 As for the second limitation, only excited states of single excitons have been experimentally observed to date \cite{Chernikov-2014-076802,Yaffe-2015-045414}, and it is straightforward to obtain excited single exciton states via grid-based numerical methods (see this work and Ref.~\cite{Berkelbach-2013-045318}). If excited trions and biexcitons have to be analyzed, one can use density matrix-based methods \cite{Berghauser-2014-125309,Ramirez-Rorres-2014-085419} or a generalized PIMC \cite{Lyubartsev-2005-6659}.

Finally, we would like to emphasize that even though ML-TMDC constitutes one of the most promising classes of 2D semiconductors at present, the results of this work are very general and could be applicable to many other materials and systems as well. For example, lead chalcogenide nanosheets are effectively 2D semiconductors with almost identical isotropic electron and hole effective masses \cite{Aerts-2014-3789,Bielewicz-2015-826}. These materials do also possess a large ``flavor multiplicity" of eight (see Sec.~\ref{sec:PIMC}) required to avoid the fermion sign problem when applying PIMC to multi-carrier bound states in their lowest energy configurations. Other potential applications include stanene \cite{Xu-2013-136804,Balendhran-2015-640} and ultrathin organic-inorganic perovskite crystals \cite{Yaffe-2015-045414}. In particular, the latter materials have been shown to have large exciton binding energies \cite{Yaffe-2015-045414} complemented by a non-hydrogenic exciton Rydberg series suggesting a non-local screening of the Coulomb interaction, Eq.~(\ref{eq:H0V0}). Such perovskite crystals could have not too different electron and hole effective masses \cite{Brivio-2014-155204,Yin-2015-8926}, rendering the numerical results of the present work directly applicable to them.

Numerical results obtained here are not applicable to phosphorene \cite{Liu-2014-4033,Li-2014-372,Prada-2015-245421,Seixas-2015-115437} due to a very large anisotropy of effective masses. However, the PIMC methodology can be straightforwardly modified to account for this. Analysis of the multi-carrier bound states in 2DS with the large anisotropy of effective masses will be the focus of our future work.

\acknowledgements

The authors are thankful to James E. Gubernatis, Timothy Berkelbach and Josiah Bjorgaard for discussions and help with the manuscript.
K. A. V. was supported by the Center for Advanced Solar Photophysics (CASP), an Energy Frontier Research Center funded by the Office of Basic Energy Sciences, Office of Science, US Department of Energy (DOE). A. S. acknowledges financial support from the NNSA of the US DOE at LANL under Contract No. DE-AC52-06NA25396


%

\end{document}